\documentclass{iucr}              

\RequirePackage{amsmath,amssymb}
\RequirePackage{textcomp}
\RequirePackage[dvips]{graphicx}

     \paperprodcode{a000000}      
     \paperref{xx9999}            
     \papertype{FA}               

     \paperlang{english}          
     \journalcode{J}              

     \journalyr{20005}
     \journaliss{1}
     \journalvol{0}
     \journalfirstpage{000}
     \journallastpage{000}
     \journalreceived{0 XXXXXXX 0000}
     \journalaccepted{0 XXXXXXX 0000}
     \journalonline{0 XXXXXXX 0000}

\graphicspath{{./figures/}}

\begin{document}                  



\title{Inferring orientation distributions in anisotropic powders of nano-layered crystallites from a single two-dimensional WAXS image}
\shorttitle{Orientation distributions of nano-stacks from a single 2D WAXS image}


\cauthor[a]{Yves}{M\'eheust}{meheust@phys.ntnu.no}{}
\author[b]{Kenneth Dahl}{Knudsen}
\author[a]{Jon Otto}{Fossum}

\aff[a]{Physics Department, Norwegian University of Science and Technology, 7491 Trondheim,  \country{Norway}}
\aff[b]{Physics Department, Institute for Energy Technology, Kjeller, \country{Norway}}


\shortauthor{M\'eheust et al.}







\maketitle                        

\begin{synopsis}
A method to determine the orientation distribution probability function of a population of nano-stacks from  the dependence of a given diffraction peak's amplitude on the azimuthal angle is proposed. It is applied to two different types of orientational order observed in systems of Na-fluorohectorite clay particles.
\end{synopsis}


\begin{abstract}
The wide-angle scattering of X-rays by anisotropic powders of nano-layered crystallites (nano-stacks) is addressed. 
Assuming that the orientation distribution probability function $f$ of the nano-stacks 
only depends on the deviation of the crystallites' orientation from a fixed reference direction, we derive a relation providing $f$ from the dependence of a given diffraction peak's amplitude on the azimuthal angle. The method is applied to two systems of Na-fluorohectorite (NaFH) clay particles, using synchrotron radiation and a WAXS setup with a two-dimensional detector. In the first system, which consists of dry-pressed NaFH samples, the orientation distribution probability function corresponds to a classical uniaxial nematic order. The second system is observed in bundles of polarized NaFH particles in silicon oil; in this case, the nanostacks have their directors on average in a plane normal to the reference direction, and $f$ is a function of the angle between a nano-stack's director and that plane. In both cases, a suitable Maier-Saupe function is obtained for the distributions, and the reference direction is determined with respect to the laboratory frame. The method only requires one scattering image. Besides, consistency can be checked by determining the orientation distribution from several diffraction peaks independently.
\end{abstract}



\section{Introduction}

The recording of a powder diffraction signal using a two-dimensional detector is a potentially simple and efficient technique for determining characteristic structural
length scales in crystals. The basis of the technique lies in that a scattering volume containing an isotropic powder consisting of a large number of particles with identical crystalline structure 
offers to the incoming beam all possible incidence orientations with respect to the crystallites. Consequently, 
all the diffraction peaks characteristic of the crystallite structure are visible in any of the azimuthal planes, and all azimuthal planes are equivalent: a simple scan of the deviation angle, $2\,\theta$, in a given azimuthal plane, provides a spectrum containing peaks at all structural length scales characteristic of the crystallites, apart from those extinct due to form factor effects. 
The scattering image collected by a two-dimensional detector  is isotropic, and its center of symmetry denotes the direction of the incident beam. Radial cuts of the images passing through this center provide identical one-dimensional spectra that can be averaged over all azimuthal angles in order to significantly increase the signal to noise ratio. 

It is well known that reconstruction of the three-dimensional crystal structure from powder data is made difficult by the overlapping of 
reflections with similar diffraction angles.
The interpretation is fairly easy in the case where only a dominant feature of the structure is investigated. A particular case is that of nano-layered systems, where the powder is made of particles, or clusters of particles, that consist of a stack of identical quasi-2D crystallites. The Bragg reflections resulting from the set of Bragg planes associated to those stacks are the most intense and can be interpreted in quite a straightforward manner. A number of natural (vermiculite, montmorillonite) and synthetic (fluorohectorite) smectite clays 
have the properties of presenting, in the weakly-hydrated states, a microstructure where particles are platelet-shaped and nano-layered. The repeating crystallite is a 1nm-thick silicate-platelet, and the stacks are held together by cations shared between adjacent platelets. The stacks can swell by intercalation of ions and molecules (water, in particular), and powder diffraction has been used extensively to study that swelling, both on natural \cite{wadaPRB90} and synthetic \cite{dasilvaPRE2002} smectite clays.
\medskip

A wide range of natural or man-made materials are aggregates of crystallites with the same crystallographic structure. 
The distribution of the crystallite orientations inside the material are rarely isotropic, hence their X-ray diffraction
signal is that of an anisotropic powder. Many examples of such polycrystalline aggregates can be found among metals. The analysis from X-ray diffraction data of the preferred crystallographic orientations in those metals, i.~e., their {\em texture}, provides information on the history of their deformation \cite{WenkBOOK85}, as it does for rocks, another type of polycrystalline materials with a larger degree of complexity (several types of crystals coexist).  This is done by decomposing the distribution of the reflections' intensities as a function of the angular direction into a superposition of distributions resulting from single reflections  ({\em pole figures}), and subsequently inferring the
orientation distribution probability (ODP) for the crystallites. 
Several general methods \cite{WenkBOOK85,BungeBOOK93} have been developed to tackle that potentially difficult (and sometimes, non-univoque) inversion. 
From the experimental point of view, 2D detectors have proved useful in such texture analyses, as they allow gathering of the data necessary to the pole figure analysis in a short time and without changing the sample position  \cite{ischiaJApplCryst2005}. See  \cite{wenkJApplCryst2003} about the practical applicability of those methods.  
Note that texture studies have recently found an interesting application in methods allowing to determine the complete 3D structure of a crystal from anisotropic powder diffraction data collected from samples with different textures \cite{lasochaJApplCryst97,wesselsScience99}.
\medskip

The study of preferential orientations in polycrystalline materials consisting of nano-layered particles (quasi-1D crystals) such as the clays presented above has made little use of the general inversion technique for texture data, since there is basically no overlap of pole figures for those materials. In the case where the ODP possesses an axial symmetry, it is classically determined experimentally by measuring a {\em rocking-curve}, in which the sample is rotated around an incidence angle $\theta$ corresponding to a well known $(00k)$ reflection, and the scattered intensity is measured at the deviation angle $2\, \theta$ as a function of $\theta$ \cite{guntertCarbon64,taylorClayMiner66}. In this paper, we present an alternative method for determining the orientation distribution from the scattering data in the case in which it only depends on one characteristic angle. When the assembly of nano-stacks  present a partial priviledged orientation, all azimuthal planes are not equivalent any more. Our method is based on monitoring the dependence of the intensity of a given reflection peak on the azimuthal angle. It can thus be carried out on a single two-dimensional wide angle scattering image. In other words, we monitor the anisotopry of that image at a given reflection. Note that monitoring of the intensity along Debye rings has previously been done in a similar manner for cubic and hexagonal materials \cite{puigmolinaZMetallkd2003,wenkJApplCryst2003}.
We do not address the influence of the ODP on the diffraction profiles  in a given azimuthal plane, a problem to which much attention has been given in the past, for systems like pyrolytic graphite \cite{guntertCarbon64} and smectite clays \cite{decourvilleJAppCryst79,planconJAppCryst80}.

The paper is organized as follows.
In section \ref{sec:theory}, we present the theoretical background of the method, for both geometries. 
The method is applied to data from two experiments corresponding to two geometrical configurations observed in systems of the synthetic smectite clay Na-fluorohectorite (in sections \ref{sec:experiments_nem} and \ref{sec:experiments_antinem}, respectively). In section \ref{sec:discussion}, we discuss the interest of the method. Section \ref{sec:conclusion} is the conclusion.

\section{Theoretical background}
\label{sec:theory}

We consider a clay sample placed at the origin $O$ of the laboratory frame $(O,\boldsymbol x,\boldsymbol y, \boldsymbol z)$ (see Fig.~1). An incident X-ray beam parallel to $\boldsymbol x$, and of wavelength $\lambda$, is scattered by the sample, and the scattered intensity recorded in a detector plane $\boldsymbol y O' \boldsymbol z$, $O'$ being the point where the direct beam would hit the detector in the absence of a beam stop behind the sample. The sample consists of a population of nano-layered crystallites consisting of $\sim 100$ stacked identical layers. The orientation distribution $f$ of the crystallite population with respect to the laboratory frame is assumed known and anisotropic, resulting in an anisotropic distribution of the scattered intensities in the detector plane.
The running point M on the detector is denoted by the direction of the line $O\boldsymbol u$ passing through the points $O$ and $M$; the corresponding angular frame consists of the scattering deviation
angle $2\, \theta$ and the azimuthal angle $\phi$.

\subsection{Diffraction by an anisotropic distribution of nano-stacks}

The orientation of each crystallite's director, $\boldsymbol n$, with respect to the fixed frame, is denoted by the angular frame  ($\Theta$,$\Phi$) (see Fig.~1),
 with $\Theta$ in the range 
 $[0;\pi/2]$ and $\Phi$ in the range $[-\pi;\pi]$. $\Theta$ is the angle between
the crystallite's director and the plane $\boldsymbol y O \boldsymbol z$, while $\Phi$ is the 
azimuthal angular coordinate of the director.

Let us first consider a single crystallite whose stacking planes make with the incident beam an angle $\theta$, corresponding to the Bragg condition for the stack. This implies that it mostly scatters along the direction $O\boldsymbol u$, i.e, that the azimuthal direction for scattering is $\phi=\Phi$, and the deviation
angle is $2\, \theta = 2\, \Theta$ (see figure~2(a)).
As the azimuth $\phi$ describes the $[-\pi;\pi]$ range, the trajectory on the unit sphere of the director for the crystallites that scatter along that azimuth describes a circle in 
the plane $x=\sin \theta$ (see Fig.~2(b)).
This director is therefore  defined in the frame $(\boldsymbol x,\boldsymbol y, \boldsymbol z)$
by the coordinates
\begin{equation}
\label{eq:n_vs_phi}
\boldsymbol n_{2\, \theta}(\phi)
\begin{pmatrix}
\sin \theta\\
- \cos \theta \: \sin \phi\\
~ ~ \cos \theta \: \cos \phi
\end{pmatrix}
\text{~ ~.}
\end{equation} 
\medskip

Next, we assume that the angular distribution $f$ only depends on an single angle $\alpha$, which is the angle between a crystallite's director $\boldsymbol n$ and a reference orientation $\boldsymbol n_0$. The distribution is normalized with respect to the total solid angle available to the description, i.~e.
\begin{equation}
\label{eq:normalisation}
\int_0^{2\,\pi} \int_0^{\pi} f(\alpha) \: \sin \alpha \: d\alpha \: d\beta = 1 \text{~ ~,}
\end{equation}
where ($\alpha$,$\beta$) is a standard spherical angular frame.
We assume that the size- and orientation- distribution of the 
particles are uncorrelated, and that the beam defines a scattering volume that is large enough for the size distribution of the scatterers to be entirely contained in it. Hence, we assume
that the amplitude of a diffraction peak in the direction ($2\, \theta$,$\phi$) is proportional to the number of scatterers that satisfy the Bragg condition for that direction, and that the proportionality constant is independent of $\phi$. This also assumes that the scattering peak is well separated from other peaks along the $2\, \theta$ line. We denote $I_{2\theta}(\phi)$ the amplitude of the diffraction peak at deviation angle $2\, \theta$ and at a given azimuthal angle value $\phi$, and $\alpha_{2\, \theta}(\phi)$ the angle between the reference director $\boldsymbol n_0$ and that of the crystallites that diffract in the $(2\, \theta, \phi)$ direction. The intensity $I_{2\theta}(\phi)\; \Delta \phi$ along a length $\Delta \phi$ of the angular scattering cone of half-opening $2\theta$ is proportional to the quantity $f(\alpha_{2\, \theta}(\phi))\: \Delta \alpha(\phi,\Delta \Phi)$, where the angular length $\Delta \alpha$ is that travelled by $\boldsymbol n$ on the unit sphere as $\phi$ describes the range $\Delta \phi$. The trajectory in Fig.~2(b) 
is travelled at constant speed when varying $\phi$ 
uniformly, as appears from the vector $d\boldsymbol n / d\phi$ having a norm $\cos \theta$ (computed from Eq.~(\ref{eq:n_vs_phi})), and thus independent of $\phi$. Therefore, 
$\Delta \alpha$ only depends on $\Delta \phi$, which yields:
\begin{equation}
\label{eq:Iofphi_to_fofalpha2}
I_{2\theta}(\phi)\: \Delta \phi \propto f(\alpha_{2\, \theta}(\phi))\: \Delta \alpha(\Delta \phi) \text{~ ~.}
\end{equation}
Consequently, a uniform sampling of the azimuthal profile of a WAXS peak in the scattering figure is directly related to the orientation distribution for the scatterers, $f$. In particular, a constant profile corresponds to a uniform angular distribution of the platelets, i.e, the scattering figure of an isotropic clay powder is isotropic, as expected.

The angle $\alpha_{2\, \theta}(\phi)$ can be computed from its cosine which equals 
$\boldsymbol n_0 \cdot \boldsymbol n_{2\, \theta}(\phi)$. The orientation with respect to the laboratory frame of the mean director $\boldsymbol n_0$ is defined by the angles $\Theta_0$ and $\Phi_0$, and thus its coordinates in the $(\boldsymbol x,\boldsymbol y, \boldsymbol z)$ frame are
\begin{equation}
\label{eq:coordinates_n0}
\boldsymbol n_0
\begin{pmatrix}
\sin \Theta_0\\
- \cos \Theta_0 \: \sin \Phi_0\\
~ ~ \cos \Theta_0 \: \cos \Phi_0
\end{pmatrix}
\text{~ ~.}
\end{equation} 
Thus,
\begin{equation}
\label{eq:alpha_vs_phi}
\alpha_{2\, \theta}(\phi) =  \text{acos} \left [ \sin \theta \, \sin \Theta_0 
+ \cos \theta \, \cos \Theta_0 \: \cos (\phi-\Phi_0 ]
\right ) \text{ ~.}
\end{equation}
\medskip
For a uniform sampling of the azimuthal scattering intensity profile $I_{2\theta}(\phi)$, one can directly relate the function $I_{2\, \theta}(\phi)$ and $f(\alpha)$ bringing together Eq.~(\ref{eq:Iofphi_to_fofalpha2}) and (\ref{eq:alpha_vs_phi}) according to
\begin{equation}
\label{eq:Iofphi_to_fofalpha3}
I_{2\, \theta}(\phi) \propto f\left ( \text{acos} \left [ \sin \theta \, \sin \Theta_0 
+ \cos \theta \, \cos \Theta_0 \, \cos (\phi- \Phi_0)  \right ] \right )\text{~ ~.}
\end{equation}

From a given functional form for the distribution $f$, with a given set of shape parameters (peak position, width, reference level), it is possible to fit a function in the form 
(\ref{eq:Iofphi_to_fofalpha3}) to the experimental data, and infer both the distribution's shape parameters and the reference orientation $\boldsymbol n_0(\Theta_0,\Phi_0)$. The obtained parameters are independent of the particular reflection ($2\, \theta$ or $q$) chosen, which provides a interesting way of checking the results' consistency.

\subsection{Two particular geometries}

\subsubsection{Uniaxial nematic configuration:}
\label{sec:def_nematic}

In the classical uniaxial nematic geometry, the reference orientation is the mean orientation for the 
population of nano-stacks. The smaller $\alpha$, the more crystallites have their orientation along $\boldsymbol n(\alpha)$. The orientation distribution $f$ is peaked in $0$ and monotically decreasing to its minimum $f(\pi/2)$.
An order parameter \cite{deGennesBOOK93_orderparam}, defined as 
\begin{equation}
S  = \frac{1}{2} \: \left \langle 3\, \cos^2 \alpha -1\right \rangle_f \text{~,}
\end{equation}
\begin{equation}
\label{eq:def_orderparam}
\text{i.e ~ ~} S  = \pi \, \int_0^{\pi} \! \! \!  ( 3 \: \cos^2 \alpha -1 )  \: f(\alpha)
\: \sin \alpha \; d\alpha\text{~,}
\end{equation}
quantifies the overall alignment of the crystallites. For $S=0$, the population is completely isotropic, while for $S=1$ all crystallites are perfectly aligned. The alignment can also be quantified in terms of the root mean square
(RMS) width $w_f$ of the distribution, according to
\begin{equation}
\label{eq:def_RMS}
w_f^2= \langle \alpha^2 \rangle_f = 2\, \pi \: \int_0^{\pi} \! \! \! \alpha^2  \: f(\alpha)   
\: \sin \alpha \: d\alpha \text{~ ~.}
\end{equation}

\subsubsection{Uniaxial "anti-nematic" configuration:}
\label{sec:def_antinematic}

In the configuration that we denote "anti-nematic", the crystallites have their directions on average in a plane 
perpendicular to the reference direction. In other words, the stacking planes of the crystallites contain the reference direction, on average. The distribution $f$ is peaked around $\pi/2$, monotically increasing with $\alpha$ from its minimum value $f(0)$. By analogy to the nematic description,
an order parameter can be defined as well, according to:
\begin{equation}
S  = \left \langle 3\, \cos^2 \left ( \frac{\pi}{2} - \alpha \right ) - 2 \right \rangle_f
= \left \langle 3\, \sin^2 \alpha - 2 \right  \rangle_f \text{~,}
\end{equation}
\begin{equation}
\label{eq:def_orderparam2}
\text{i.e~ ~}S  = 2\, \pi \, \int_0^{\pi} \! \! \! \left ( 3 \: \sin^2 \alpha  - 2 \right ) \: f(\alpha) 
\: \sin \alpha \; d\alpha\text{~,}
\end{equation}
and an RMS angular width according to:
\begin{equation}
w_f^2  = \left \langle \left ( \frac{\pi}{2} -  \alpha \right ) ^2 \right \rangle_f  \text{~,}
\end{equation}
\begin{equation}
\label{eq:def_RMS2}
\text{i.e ~ ~}  w_f^2   =  2\, \pi \: \int_0^{\pi} \! \! \! 
\left ( \frac{\pi}{2} - \alpha \right ) ^2  \: f(\alpha)  \: \sin \alpha \: d\alpha  \text{~ ~.}
\end{equation}

\subsection{Modelled azimuthal profiles}
\label{sec:modelled_profiles}

Fig.~3 shows expected azimuthal profiles, computed from Eq.~(\ref{eq:Iofphi_to_fofalpha3}), in the case of a 
nematic geometry where the distribution $f$ is of the classical Maier-Saupe form observed in uniaxial nematic liquid 
crystals \cite{MaierNatForsch58,MaierNatForsch59}:
\begin{equation}
\label{eq:MSfunction}
f(\alpha) \propto \exp(m\, \cos^2 \alpha)\text{ ~.}
\end{equation}
For a reflection corresponding to a particular value of the deviation angle $2\, \theta = 3.29${\textdegree}, two values of the distribution's angular width $w_f$ ($15$ and $30${\textdegree}), and three different orientations of the mean director $\boldsymbol n_0$ ($\Theta_0 = -5, 0$ and $5$\textdegree) have been used. The profiles are normalized to an integral value $1$ over the whole azimuthal range. A profile corresponding to $\Theta_0 = 5${\textdegree} for $2\, \theta=6.54$ is also shown.

Fig.~3 shows clearly how the parameters describing the orientational ordering of the crystallites translate into the azimuthal dependence of a diffraction peak's intensity. The angular width of the orientation distribution controls the width of the profile's peaks. 
The azimuthal coordinate of the mean director, $\Phi_0$, only has the effect of 
translating the profile along the $\phi$ scale, while $\Theta_0$ is responsible
for the curve's asymmetry with respect to the azimuth $\phi = \Phi_0 + 90^\circ$, or, equivalently, the asymmetry of the 2D diffractogram with respect to the plane $\boldsymbol x O \boldsymbol y$ (see 
Fig.~1).
In addition, the magnitude of this asymmetry depends on the value of $2\, \theta$: if there is asymmetry (i.e, if $\Theta_0 \neq 0$), it shall
be more pronounced for the higher order peaks of a given reflection.

\section{Application to a nematic geometry}
\label{sec:experiments_nem}

\subsection{Fluorohectorite, a synthetic smectite clay}

Fluorohectorite belongs to the family of smectite (or swelling 2:1) clays, whose 
base structural unit is 1nm-thick phyllosilicate platelet consisting of two inverted silicate tetrahedral sheets that share their apical oxygens with a tetrahedral sheet sandwitched in between \cite{veldeBOOK95}. 
Fluorohectorite has a chemical formula per unit cell X$_x$-(Mg$_{3-x}$Li$_x$)Si$_4$O$_{10}$F$_2$.
Substitutions of Li$^{+}$ for Mg$^{2+}$ in part of the fully occupied octahedral 
sheet sites are responsible for a negative surface charge along the platelets. This results here in a large surface charge, $1.2$ e$^-$\cite{kaviratnaJPhysChemSol96} (as opposed to $0.4$ e$^-$ for the synthetic smectite clay laponite), allowing the platelets to stack by sharing an intercalated cation X, for example Na$^{+}$, Ni$^{2+}$, Fe$^{3+}$. The resulting nano-stacks remain stable in suspension, even at low ionic strength of the solvent, and are therefore present both in water solutions, gels \cite{diMasiPRE2001} and weakly-hydrated samples \cite{knudsenPHYSICA2004}.

Wide-angle X-ray scattering of gel \cite{diMasiPRE2001} and weakly-hydrated \cite{dasilvaPRE2002} samples of Na-fluorohectorite (for which X=Na$^{+}$), have shown that the characteristic platelet separation is well defined, and varies stepwise as a function of the surrounding temperature and humidity \cite{dasilvaPRE2003}. Samples are anisotropic, but the width of the corresponding WAXS peaks allows inferring the typical nano-stack thickness, which is of about $80-100$ platelets. This thickness is significantly larger than thicknesses ($\lesssim 20$ platelets) for which asymmetry and irrationality in the $(00k)$ reflections are known to occur \cite{dritsBOOK90_thinstacks}, which makes the analysis of $(00k)$ peaks in powder diffraction spectra of Na-fluorohectorite relatively easy.

\subsection{Dry-pressed samples of Na-fluorohectorite}

Dry-pressed samples of Na-fluorohectorite were prepared from raw fluorohectorite powder purchased from Corning Inc. (New York). The raw powder was carefully cation-exchanged so as to force the replacement of all intercalated cations by Na$^{+}$ \cite{lovollPHYSICAB2005}. The resulting wet gel was then suspended in deionized water again, and after sufficient stirring time, water was expelled by applying a uniaxial load in a hydrostatic press.
The resulting dry assembly was then cut in strips of 4mm by 47mm.

The resulting samples are anisotropic, with a marked average alignment of the clay crystallites parallel to the horizontal plane \cite{knudsenPHYSICA2004}, as shown in the two-dimensional sketch of the sample geometry, in the insert to Fig.~4.
A nematic description such as that presented in section \ref{sec:def_nematic}, with
a mean director $\boldsymbol n_0$ close to the direction of the applied uniaxial load, describes their 
geometry well.

\subsection{Experimental setup}
\label{sec:setup1}

Two-dimensional diffractograms of dry-pressed samples were obtained on beamline BM01A at the European Synchrotron Radiation Facility (ESRF) in Grenoble (France). We used a focused beam with energy $17.46$ keV corresponding to a 
wavelength $0.71$\AA. The scattering images were recorded using a two-dimensional MAR345 detector, with a resolution of $2300$ pixels
across the diameter of the $345$mm wide circular detector. The distance between the sample and the detector was calibrated to $375.25$mm using a silicon powder standard sample (NBS640b).

Fig.~4 shows a simplified sketch of the experimental geometry.
The samples were lying with the average nano-stack director close to vertical.
A sample cell allowing humidity- and temperature- control in the surrounding atmosphere is not shown.
A more complete description of the sample environment can be found in \cite{meheustPRE2005}.

\subsection{Image analysis}

The center part of a two-dimensional scattering image obtained from a dry-pressed Na-fluorohectorite sample at a temperature of $98${\textcelsius} is shown in Fig.~5(a).
 The image anisotropy appears clearly, with 
a much stronger intensity of some of the diffraction rings along the vertical direction. Some grainy-looking isotropic rings can also be seen, they are due to impurities (mostly mica). In what follows, we infer the orientation distribution for the clay crystallites from three rings, separately. These rings correspond to the three first orders
of the same reflection.

\subsubsection{Method:}
The images were analyzed in the following way. After finding the image center (incident beam direction), we compute diffraction lines at azimuthal angle values $\phi_m$, varying $\phi_m$ between $0$ and $359${\textdegree} by steps of $1${\textdegree}. For a given value of $\phi_m$, the diffraction line is obtained by integrating over $\phi$ the image subpart corresponding to values of $\phi$ in the range $[\phi_m-\delta \phi/2; \phi_m+\delta \phi/2]$, where $\delta \phi=5$\textdegree. Six of these powder diffraction profiles are shown in 
Fig.~6.
They exhibit $6$ peaks characteristic of the nano-layered structure of the crystallites: the $3$ first orders for the population of crystallites with no water layer intercalated (0WL), and the orders 1, 2, and 4 for the population of crystallites with one water layer intercalated. At this temperature, 
the 1WL state is unstable \cite{dasilvaPRE2003}; it is present in these pictures because the transition between the two hydration states is not finished yet. The analysis is carried out on the peaks 0WL(001), 0WL(002) and 0WL(003), which are the most intense, and are weakly perturbed by other peaks. The corresponding characteristic length scales are listed in table \ref{tab:dvalues}; they are in good agreement with previous measurements \cite{dasilvaPRE2002}.

For each of the three peaks 0WL(001), 0WL(002) and 0WL(003), the peak's intensity was plotted as a function of $\phi$, and the obtained profiles were fitted with Eq.~(\ref{eq:Iofphi_to_fofalpha3}). 
The three normalized profiles are shown in Fig.~7;
they have been translated vertically for clarity. As a functional form for the orientation distribution, we chose the Maier-Saupe function (Eq.(~\ref{eq:MSfunction})). We also allowed for a constant background $I_0$ to be accounted for. This  resulted in fitting to the profiles a function in the form
(using Eq.~(\ref{eq:Iofphi_to_fofalpha3}) and (\ref{eq:MSfunction}))
\begin{equation}
\label{eq:fitting_function}
I_0 + C \: \exp \left [m\,\left (  \sin \theta \, \sin \Theta_0 
+ \cos \theta \, \cos \Theta_0 \: \cos (\phi-\Phi_0)  \right )^2 \right ] \text{~ ~,}
\end{equation}
with the fitting parameters $I_0$, $C$, $m$, $\Theta_0$ and $\Phi_0$. 
Note that Eq.~(\ref{eq:fitting_function}) depends on the diffraction peak chosen, as the value for the half-deviation angle 
$\theta$ is no parameter for the fit but is different for each of the three profiles. 
As each of the fit parameters controls a particular shape aspect of the profiles (see section \ref{sec:modelled_profiles}), we can derive estimates for the parameters $I_0$, $C$, $m$, $\Theta_0$ and $\Phi_0$, using a few approximations (see appendix~\ref{sec:initial_parameters}), and use these estimates as initial parameters for the fits. The fitting procedure is fast and the fitted parameters do not depend on the initial guesses. 

Using the fitted model, we then derived the quantities characteristic of the orientational ordering in the samples, namely the order parameter $S$ (Eq.~(\ref{eq:def_orderparam})) and the angular RMS width $w_f$ (Eq.~(\ref{eq:def_RMS})).

\subsubsection{Results:}
\label{sec:results1}

The fits are shown in Fig.~7
as solid lines, on top of the data points.
As $\phi$ moves away from the peak values $\Phi_0$ and $\Phi_0+\pi/2$, and as the intensity goes down to the noise level of the diffraction profiles shown in Fig.~6, 
the significance of the intensity $I(\phi)$ in terms of orientation distribution of the profiles is somewhat diluted.
Hence, the points in the flat sections outside the peaks in Fig.~7
are little relevant when fitting the model to the data. We therefore estimate an uncertainty on the fitted parameters from the fluctuations in the fit parameters obtained by varying the range of data points (between $30$ and $90$ degrees from the peaks) considered for the fit. This uncertainty is considerably larger than the statistical error on the fitted parameters themselves, for a given fit.

For each of the profiles, the fitted parameters and the derived quantities are shown in table~\ref{tab:fit_results}.
The results obtained from the different orders are consistent with each other within the uncertainties, except for the parameter $\Theta_0$ where a $0.5$\textdegree discrepancy exists. This is however small, and the first order profile was
expected to behave slightly differently from the others as the corresponding peak is not totally separated from the 1WL(001) peaks (see Fig.~6). This profile also gives a value slightly different for the angular width, but consistent 
with the other orders within the error bars. We can thus estimate the average orientation of the nano-stack population
to $(\Theta_0,\Phi_0)=(5.8$\textdegree$\pm0.5$\textdegree$;176.3$\textdegree$\pm0.1$\textdegree). This
value results from a possible non-parallelism of the average crystallite orientation from the direction of the constraint applied during sample preparation, and from a possible misaligment of the sample (a few degrees at most) from the horizontal direction.

\section{Application to an "anti-nematic" configuration}
\label{sec:experiments_antinem}

We now apply the technique to data obtained in wide angle X-ray scattering experiments from bundles of Na-fluorohectorite particle in silicon oil. 

\subsection{Electorheological chains of fluorohectorite particles}

Fluorohectorite samples were prepared by suspending $1.5$ \% by weight of clay particles in silicon oil (Rothiterm M150, with a viscosity of 100cSt at 25\textcelsius). The hydration state of the particles prior to suspension in the oil was 1WL.
When an electric field above $500$ V.mm$^{-1}$ was applied, the suspended crystallites polarize, and the subsequent induced dipoles interact with each other, leading to aggregation of the clay crystallites in columnar bundles \cite{fossumPRL2005}. This can be seen in the microscopy picture shown in Fig.~8.
These bundles are on average parallel to the applied electric field. 

\subsection{Experimental setup}

The orientation of the polarized crystallites inside the bundles was investigated using synchrotron wide angle X-ray scattering. A top view sketch of the experimental setup is shown in Fig.~9:
the suspensions were contained in a sample cell in between two vertical copper electrodes separated by a $2$mm gap. 
A $1$ kV electric field was applied between the electrodes. Horizontal bundles of polarized particles were observed to form within $20$s after the field was switched on, and two-dimensional diffractograms were recorded after formation of the bundles. The experiments were performed at the Swiss-Norwegian Beamlines at ESRF, using the same beam energy and detector as presented in section~\ref{sec:setup1}.

\subsection{Image analysis}

Fig.~5(b) displays the center of a scattering image obtained after the chains have settled in the electric field. This WAXS data contains an significant background, with a broad ring due to diffuse scattering from the oil, but the 1WL$(001)$ peak characteristic of the nano-layered structure of the crystallites is clearly visible. It is observed to be much more intense along the horizontal direction, which means that the directors of the crystallites are on average in a plane perpendicular to the direction of the bundle \cite{fossumPRL2005}. Furthermore, 
an axial symmetry around the horizontal direction perpendicular to the X-ray beam, that is, around the average direction for the bundles, is expected. Thus, the antinematic description presented in paragraph~\ref{sec:def_antinematic} is well suited to describe the orientational ordering in the chains.

\subsubsection{Method:}
The image analysis was done in the exact same way as for the nematic geometry studied previously.
It is carried out on the ring corresponding to the first order reflection for the 1WL hydration state, as shown by the arrow in Fig.~5(b).
The choice of a Maier-Saupe functional form for the orientation distribution is not only phenomenological here. Indeed, this precise shape is expected in a system of identical induced dipoles where alignment by an external field competes with the homogeneizing entropy \cite{fossumPRL2005,meheustPRE2004c}.

\subsubsection{Results:}
The normalized azimuthal profile extracted from the two-dimensional scattering image is shown in 
Fig.~10, with the corresponding model fitted on the whole range of azimuthal angles. Despite a significant noise away from the peak positions, the fit is satisfying. The uncertainties on the angular RMS width and order parameters (see table~\ref{tab:fit_results}, column marked ER) are estimated in the way described in paragraph \ref{sec:results1}.

\section{Discussion}
\label{sec:discussion}

The method reported in this paper displays several strong points with respect to a technique based on a one-dimensional recording, like the "rocking curve" technique. The first one is that it requires very little data acquisition time, since only one recording is necessary. Such a recording can be obtained in a couple of minutes at a 3rd generation synchrotron source. It therefore allows in situ measurements of the orientation distribution while changing 
experimental parameters. In the two examples given here, it could be the surrounding humidity and the magnitude of the applied electric field, respectively. The second benefit is that it allows determination of the reference orientation in the three dimensions of space. The rocking curve method provides the inclination $\Theta_0$ of that direction with respect to planes normal to the incident beam in the case where $\Phi_0=0$ only. It does not intrinsically provide correct estimates of the angles $\Theta_0$ and $\Phi_0$ without prior time-consuming adjustment of the sample azimuthal orientation so as to ensure the condition $\Phi_0 = 0$. Besides, it introduces a bias in the estimated width of the orientation distribution in the case where $\Phi_0 \neq0$.

We determine an uncertainty on the inferred ordering parameters from the dispersion on the fitted values when changing the range of azimuth angles used for the fit. This uncertainty is therefore dependent on the noise level of the scattering image, and the process is more trustworthy if the diffraction peak considered is more intense. How the accuracy of the method depends on the peak's intensity with respect to the noise level is not quantified here. How the vicinity of another peak to the peak used to determine $f$, as in the 0WL$(001)$ peak in section~\ref{sec:experiments_nem}, modifies the azimuthal profiles, is not known quantitatively either, and we propose our method as accurate when applied to a reflection whose peak is well separated from other peaks. Despite those shortcomings, we think the method has several advantages over the 
rocking curve method and allows to better constrain the description of the orientational order in the samples, because (i) the reference direction $\boldsymbol n$ is determined, (ii) no bias in the angular width is introduced in the case where $\boldsymbol n$ does not coincide with the $O\boldsymbol z$ direction of the laboratory frame, (iii) an uncertainty is determined during the fitting procedure, for each azimuthal profile considered, and (iv) measurement consistency can be checked by comparing the results obtained from several different azimuthal profiles corresponding to several different diffraction peaks, on the same single recording. Finally, the method is easily carried out practically, since the only mathematical procedure involved is the fitting of Eq.~\ref{eq:Iofphi_to_fofalpha3} to azimuthal profiles.


\section{Conclusion and prospects}
\label{sec:conclusion}

We have developed a method to determine orientation distributions in anisotropic powders of nano-layered crystallites from a single two-dimensional scattering image, in the case when it only depends on the deviation from a reference direction. The method relies on fitting the proper relation to the azimuthal decay of a given diffraction peak's amplitude. It allows 
determination of the reference direction and of the angular dispersion of the distribution.
It was applied succesfully to data obtained from two synchrotron X-ray scattering experiments, on two different systems of synthetic smectite clay particles where the particle assemblies exhibit a nematic and anti-nematic ordering, respectively. In the first case, the consistency of the inferred quantities was checked by comparing the results obtained from three different orders of the same reflection.

The method is promising as it is performed on a single two-dimensional recording, and therefore allows in situ determination of orientational ordering in samples when thermodynamical conditions are varying. It shall be used in this respect, in the future.

\appendix
\section{Initial parameters for the fits}
\label{sec:initial_parameters}

The initial parameters for the fitting function in the form (\ref{eq:fitting_function}) are set to 
prior estimates of those parameters. The prior estimates are obtained 
from the data profiles using a few approximations, as explained below.

The parameter $I_0^{(i)}$ is first estimated as the minimum value of the data plot. A running average is performed on the data in order to smooth out the noise and obtain a curve with only two maxima: those corresponding to the peaks. The value for $\Phi_0^{(i)}$ is set to the azimuthal position of the largest peak. Its peak intensity, $I_1$ and that of the second peak, $I_2$ are determined. The half-width at half maximum $w$ of the largest peak is estimated numerically. The parameter $m^{(i)}$ is then defined from $w$ by stating:
\begin{equation}
\exp( m^{(i)} \: \cos^2 w ) = \frac{\exp(m^{(i)})}{2} \text{~,}
\end{equation}
which provides the relation
\begin{equation}
m^{(i)} = \frac{\ln 2 }{1- \cos^2 w} \text{ ~.}
\end{equation}
The maxima  $I_1$ and $I_2$ are obtained for $\phi=\Phi_0$ and $\phi=\Phi_0+\pi$, hence one can state:
\begin{equation}
\label{eq:getting_Theta_0}
\left \{ 
\begin{aligned}
I_1 - I_0^{(i)} & = C\: \exp \left (m^{(i)} \, \cos ^2 (\theta - \Theta_0^{(i)}) \right )\\
I_2 - I_0^{(i)} & = C\: \exp \left (m^{(i)} \, \cos ^2 (\theta + \Theta_0^{(i)}) \right )\\
\end{aligned}
\right .
\text{ ~ .}
\end{equation}
From this, it follows that 
\begin{equation}
\ln \left ( \frac{I_1 - I_0^{(i)}}{I_2 - I_0^{(i)}} \right ) = m^{(i)} \: \sin (2 \, \theta) \: \sin (2\, \Theta_0^{(i)})\text{ ~.}
\end{equation}
from which an estimate of $\Theta_0$ is obtained:
\begin{equation}
\Theta_0^{(i)} = \frac{1}{2} \: \text{Asin} \left [ \frac{1}{m^{(i)}\, \sin (2\, \theta)} \: \ln \left ( \frac{I_1 - y_0^{(i)}}{I_2 - y_0^{(i)}} \right ) \right ] \text{ ~.}
\end{equation}
Finally, going back to (\ref{eq:getting_Theta_0}), we obtain $C^{(i)}$ in the form
\begin{equation}
C^{(i)} = \frac{I_1 - y_0^{(i)}}{\exp \left (m^{(i)}\, \cos^2 (\theta- \Theta_0^{(i)}) \right )} \text{~.}
\end{equation}



\ack{The staff at the Swiss-Norwegian Beamlines at ESRF is gratefully acknowledged for its support during the synchrotron experiment. This work was supported by the Norwegian Research Council (NRC), through the research grants numbers 152426/431, 154059/420 and 148865/432}.
\medskip





\newpage

\begin{table}
\caption{
\label{tab:dvalues}
Characteristic momentum transfer (in nm$^{-1}$) and length scales (in \AA) measured for the 0WL diffraction peaks from dry-pressed samples of Na-fluorohectorite. "Dry" denotes the dry-pressed Na-fluorohectorite samples, while "ER" denotes the electro-rheological bundles of Na-fluorohectorite particles.}
\begin{tabular}{c|rrr}    
 Reflection & 0WL$(001)$ & OWL$(002)$ & OWL$(003)$\\
$q$ & $6.50 \pm 0.03$ & $13.02 \pm 0.03$ & $19.57 \pm 0.03$\\
$d$ & $9.67 \pm 0.04$ & $9.65 \pm 0.03$ & $9.63 \pm 0.03$\\
\end{tabular}
\end{table}

\begin{table}
\caption{
\label{tab:fit_results}
Values obtained for the fit parameters $m$, $\Theta_0$ and $\Phi_0$, and for the angular width $w_f$ and 
order parameter $S$.}
\begin{tabular}{c|rrr|r}      
Profile  & Dry $(001)$ & Dry $(002)$ & Dry $(003)$ & ER $(001)$\\
$m$ & $10.45\pm0.06$ & $10.21\pm0.27$ & $10.21\pm0.13$ &  $3.10\pm0.23$ \\
$w_f$ & $19.10\pm0.07$ & $19.38\pm0.33$ & $19.37\pm0.16$ & $24.48\pm0.50$ \\
$S$ & $0.85\pm0.01$ & $0.85\pm0.01$ & $0.84\pm0.01$ & $0.56\pm0.02$\\
$\Theta_0$ & $6.12\pm0.04$ & $5.57\pm0.10$ & $5.87\pm0.05$ & $-1.64\pm0.05$ \\
$\Phi_0$ & $176.37\pm0.01$ & $176.32\pm0.01$ & $176.37\pm0.02$ & $267.74\pm0.02$ \\
\end{tabular}
\end{table}


\begin{figure}
\includegraphics[clip=true,width=0.6\textwidth]{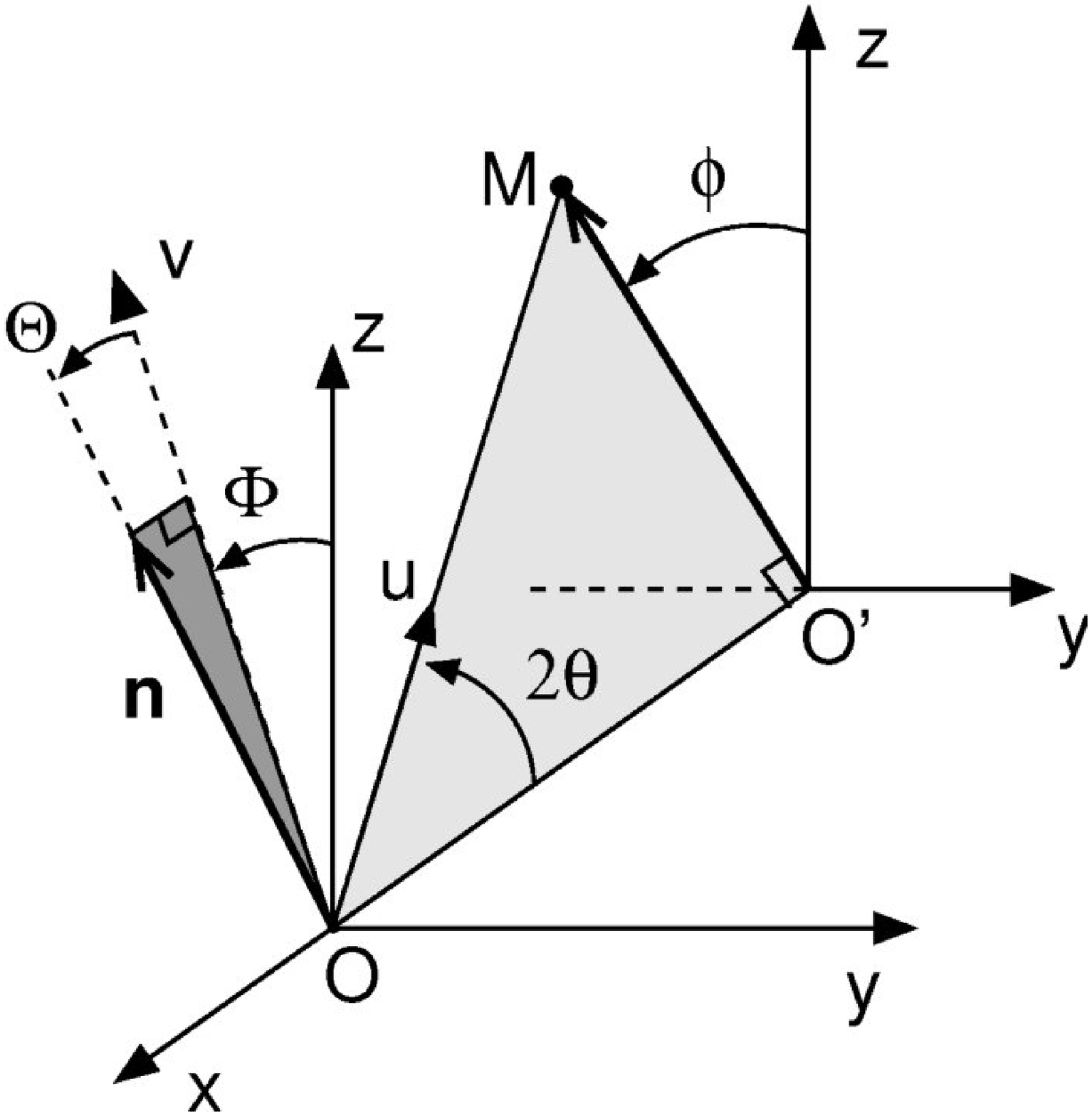}
\caption{
\label{fig:angular_frames}
In the fixed frame $(\boldsymbol x,\boldsymbol y,\boldsymbol z)$ attached to the laboratory (with $\boldsymbol x$ parallel to the incident beam), the angular direction of the diffracted beam, denoted by point $M$ in the detector plane $\boldsymbol y O' \boldsymbol z$, is defined by the deviation angle $2\, \theta$ and the azimuthal angle $\phi$. The area painted in light gray denotes the plane containing the incident and diffracted beams; it is normal to the $\boldsymbol y O \boldsymbol z$ plane and makes an angle $\phi$ with the $\boldsymbol z O \boldsymbol x$ plane. The area painted in dark gray also denotes a plane normal to the to the $\boldsymbol y O \boldsymbol z$ plane and that makes an angle $\Phi$ with the $\boldsymbol z O \boldsymbol x$ plane. It is used to define the orientation of the director $\boldsymbol n$ of a given crystallite with respect to the fixed frame, according to the angles $\Theta$ and $\Phi$.
}
\end{figure}

\begin{figure}
\caption{
\label{fig:specular}
(a) The condition of specular reflection on the Bragg planes requests that $\Theta=\theta$, as sketched here in the $\boldsymbol x O \boldsymbol u$ plane. (b) The thick curve is that described on the unit sphere by the end of vector $\boldsymbol n$ (director of the scatterer) as the azimuthal position of the diffracted beam, $\phi$, varies between $0$ and $\pi/2$. Here the deviation angle $2\, \theta$  corresponds to the Bragg reflexion (002) on a nano-layered Na-Fluorohectorite particle in the hydration state 1WL, with $\lambda=0.71$ \AA.
}
\begin{flushleft}
\vspace*{6mm}
(a)\hspace*{0.42\textwidth}(b)\\
\end{flushleft}
\vspace*{-10mm}
\includegraphics[clip=true,width=0.45\textwidth]{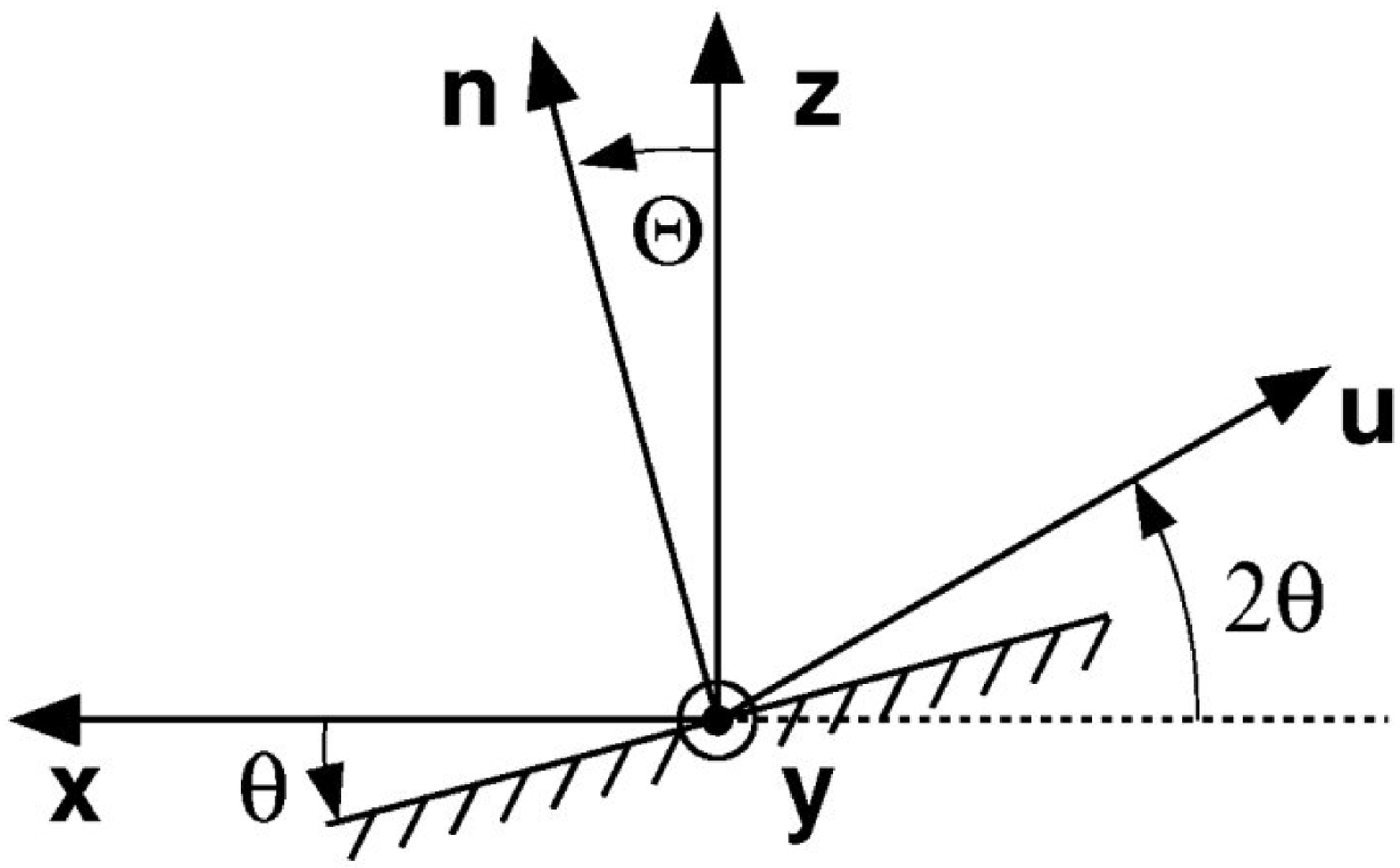}
\hfill
\hspace*{2mm} 
\includegraphics[clip=true,width=0.5\textwidth]{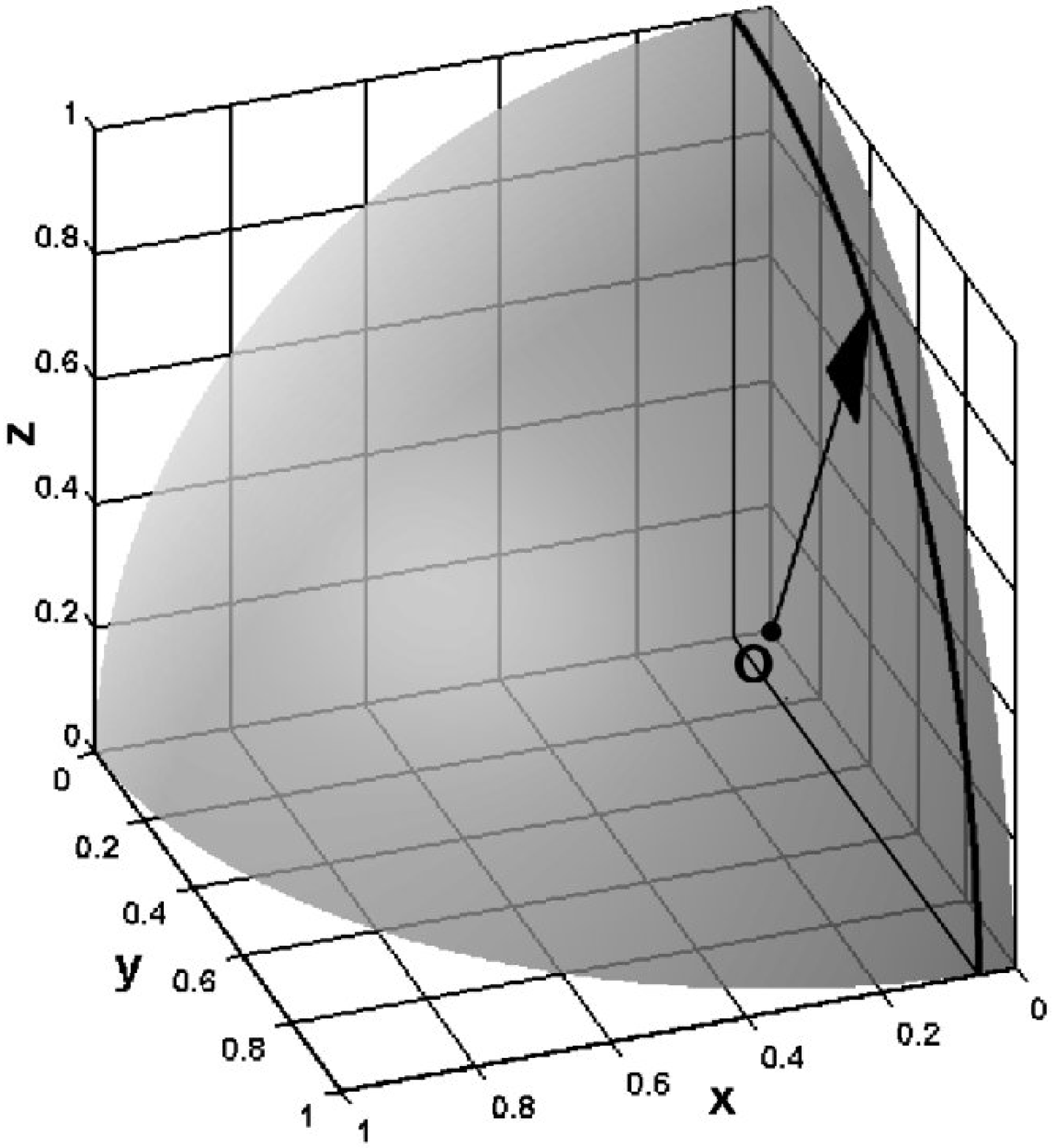}
\end{figure}

\begin{figure}
\caption{
\label{fig:azimproffunc_various_parameters}
Computed profiles describing how the intensity of a given diffraction peak (i,e, at a given deviation angle $2\, \theta$) depends on the azimuthal angle for a population of crystallites with a nematic orientational ordering. The computation was done for various values of the orientation $(\Theta_0,\Phi_0)$ of the main director with respect to the fixed frame of the laboratory, and of the 2D RMS width, $w_f$, of the nematic angular distribution  of the scatterers' directors.
Two values of the deviation angle, $2\, \theta = 6.58$\textdegree and $3.29$\textdegree, were used, which means that two diffraction peaks corresponding to two different orders of the same reflection, were used.}
\includegraphics[clip=true,width=0.98\textwidth]{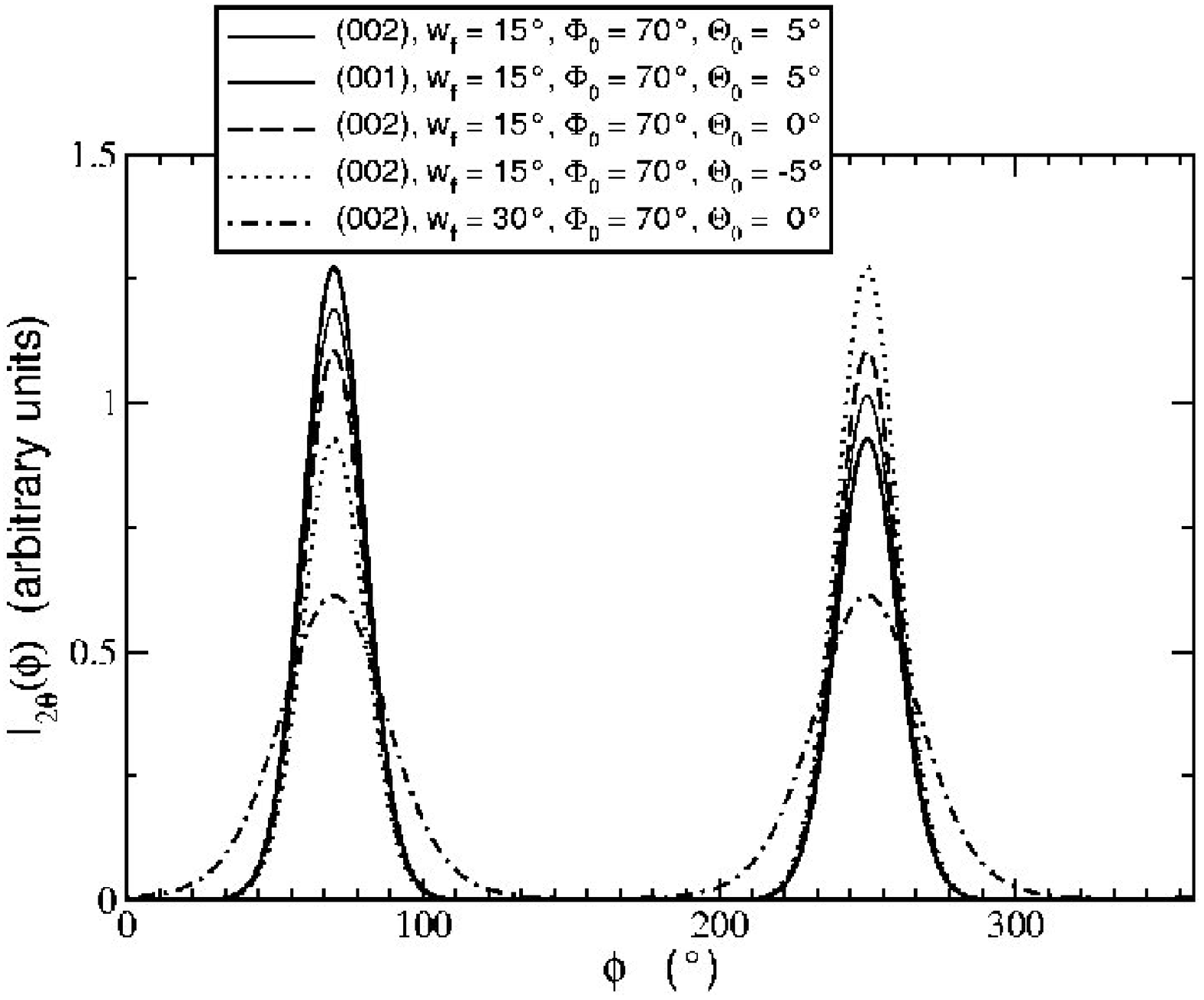}
\end{figure}

\begin{figure}
\caption{
\label{fig:setup1}
Side view sketch of the experimental scattering geometry used to scatter X-ray by dry-pressed fluorohectorite samples. The insert shows a two-dimensional cut of the sample with the nano-layered crystallites lying on average aligned along a plane close to horizontal.
}
\includegraphics[clip=true,width=0.90\textwidth]{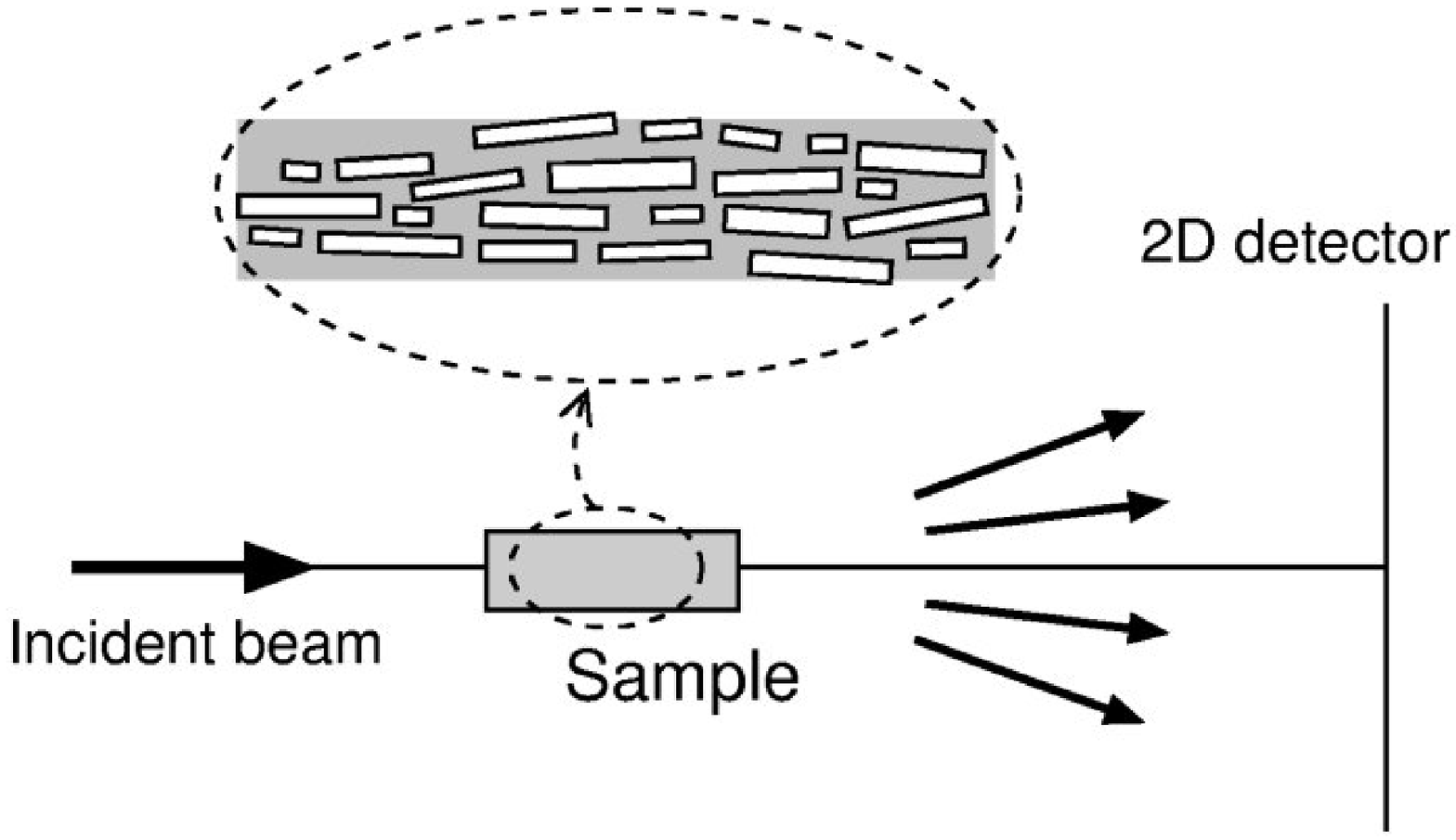}
\end{figure}

\begin{figure}
\caption{
\label{fig:2Ddiffractograms}
Center part of two-dimensional scattering images obtained from the anisotropic powders of clay nano-stacks. The anisotropic rings used to infer the orientation distributions of the crystallites are denoted by arrows. (a) Picture obtained from dry-pressed samples. The camera length was $375$ mm. (b) Picture from electrorheological chains
of crystallites in oil. The camera length was $376$ mm. The anisotropic ring in the center of (b) is the $(001)$ ring for the nano-stack structure in the 1WL hydration state. Its anisotropy is orthogonal to that observed in (a). The image magnification is $1.5$ with respect to (a). The broad peak outside the 
$(001)$ ring is due to diffuse scattering from the silicon oil.
}
(a)\includegraphics[clip=true,width=0.462\textwidth,height=0.5\textwidth]{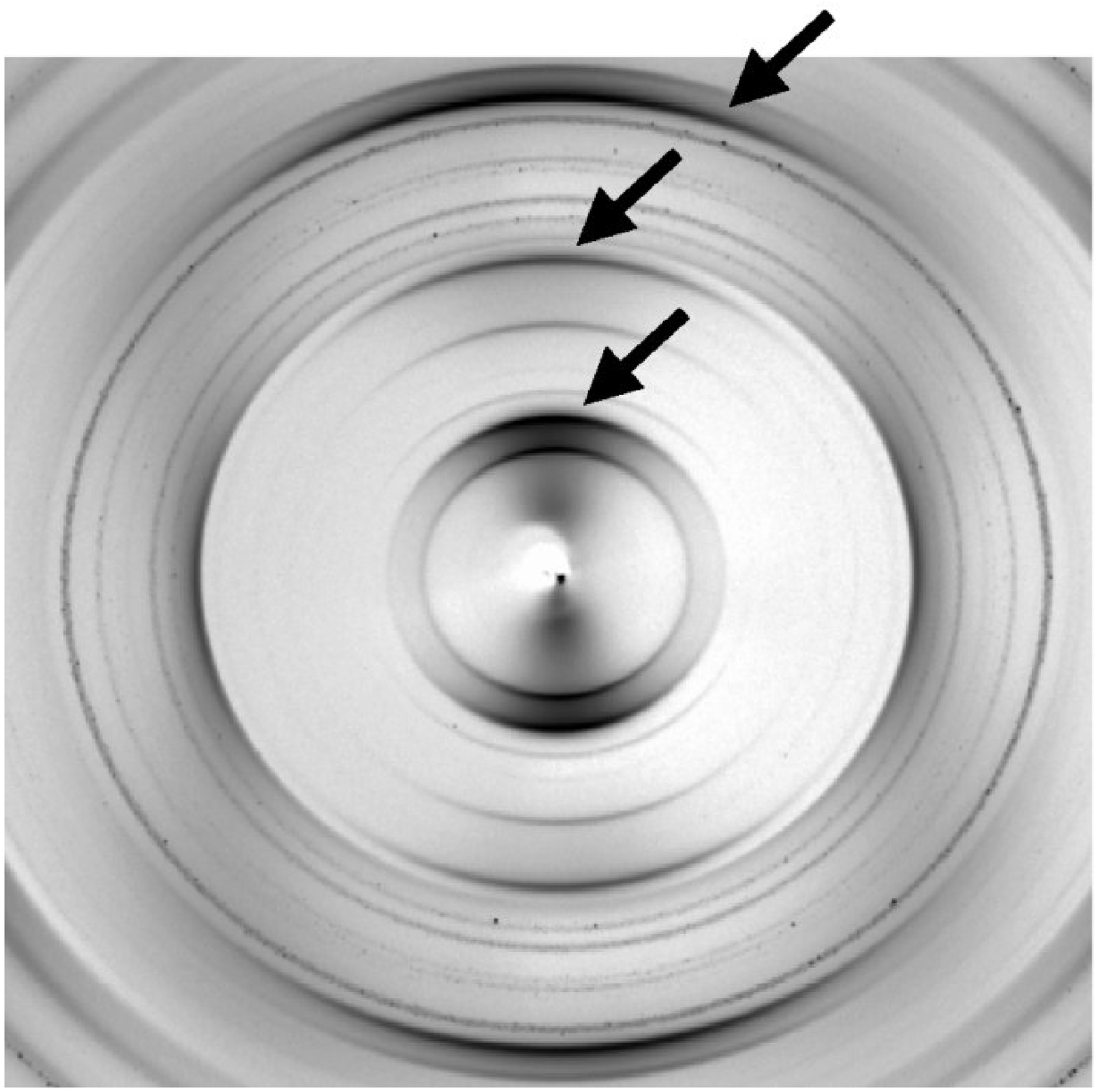}\hfill
(b)\includegraphics[clip=true,width=0.4\textwidth]{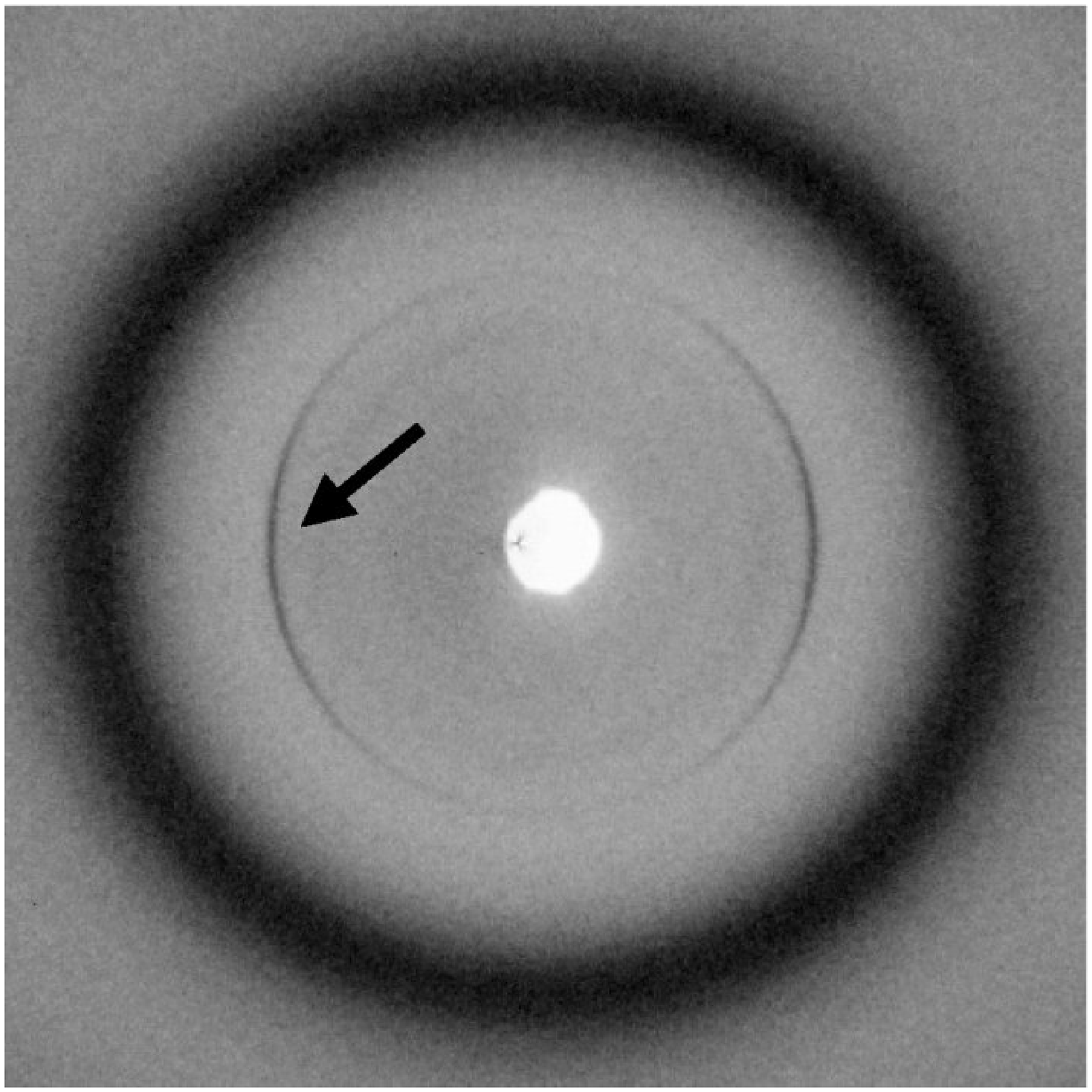}
\end{figure}

\begin{figure}
\caption{
\label{fig:NaFH_Lutnes_parallell_downT_001_variousprofiles_vsQ}
Powder diffraction profiles recorded along different azimuthal directions on the 2D detector, with a dry-pressed sample at $T=97.7${\textcelsius}. Several peaks characteristic of the nano-layered structure of the crystallites are visible. Their intensity decreases dramatatically as $\phi$ increases due to the dropdown of the number of crystallites that meet the Bragg condition at these angles.
}
\includegraphics[clip=true,width=0.95\textwidth]{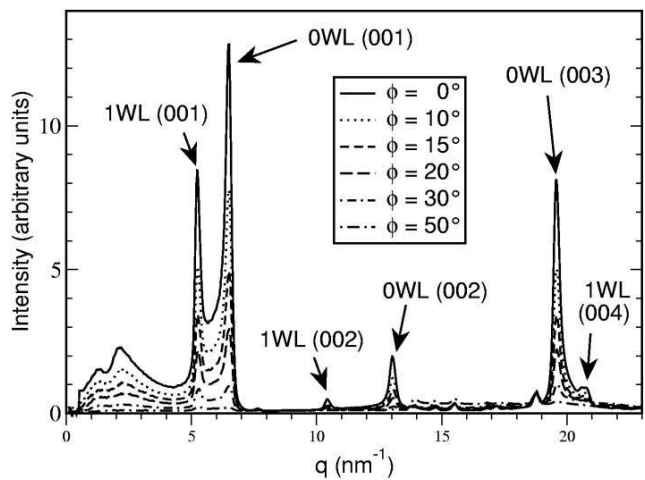}
\end{figure}

\begin{figure}
\caption{
\label{fig:azimprof_dry}
Azimuthal profiles observed from the scattering of dry-pressed samples at three values of $q$ corresponding to the three first orders of reflection by the population of crystallites in the 1WL hydration state. The profiles have been normalized and translated vertically for clarity. The fits according to Eq.~(\ref{eq:fitting_function}) are shown in solid lines on the symbol plots for the data.
}
\includegraphics[clip=true,width=0.95\textwidth]{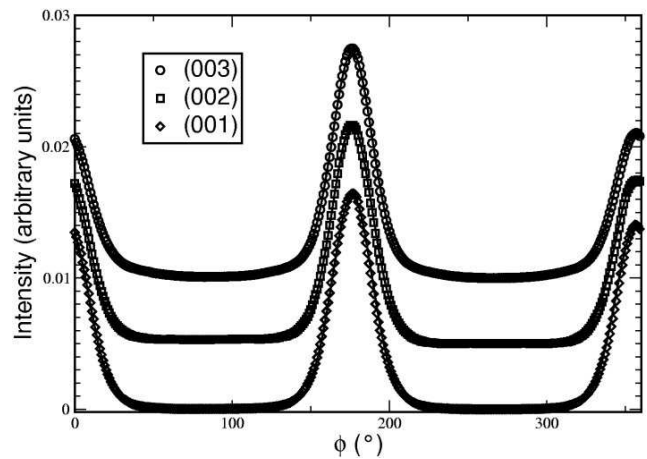}
\end{figure}

\begin{figure}
\caption{
\label{fig:ERchains}
Close view of a chain formed under the application of a strong electric field by Na-fluorohectorite crystallites suspended in silicon oil. The original color image has been filtered. The gap between the electrodes, whose boundaries are shown by two vertical bold lines on the sides of the picture, is $2$ mm.
}
\includegraphics[clip=true,width=0.995\textwidth]{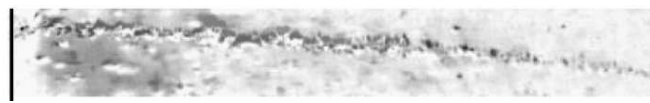}
\end{figure}

\begin{figure}
\caption{
\label{fig:setup2}
Top view sketch of the experimental scattering geometry used to scatter X-ray by electrorheological chains of Na-fluorohectorite crystallites. The electrodes are vertical, and the chains of aggregated
crystallites along a direction which is close to perpendicular to the electrodes. The disorder in the chains is exaggerated in this sketch.
}
\includegraphics[clip=true,width=0.90\textwidth]{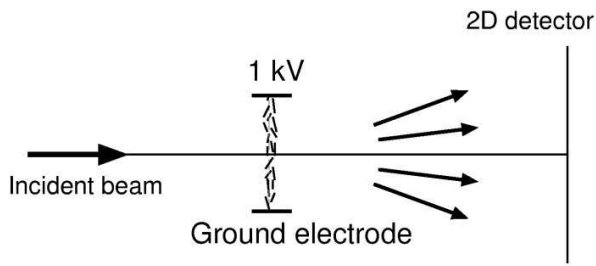}
\end{figure}

\begin{figure}
\caption{
\label{fig:azimprof_ER}
Azimuthal profile at $q=5.10$ nm$^{-1}$, corresponding to the $(001)$ peak of Na-fluorohectorite in the 1WL hydration state. The corresponding fit is plotted as a solid line on top of the data.
}
\includegraphics[clip=true,width=0.90\textwidth]{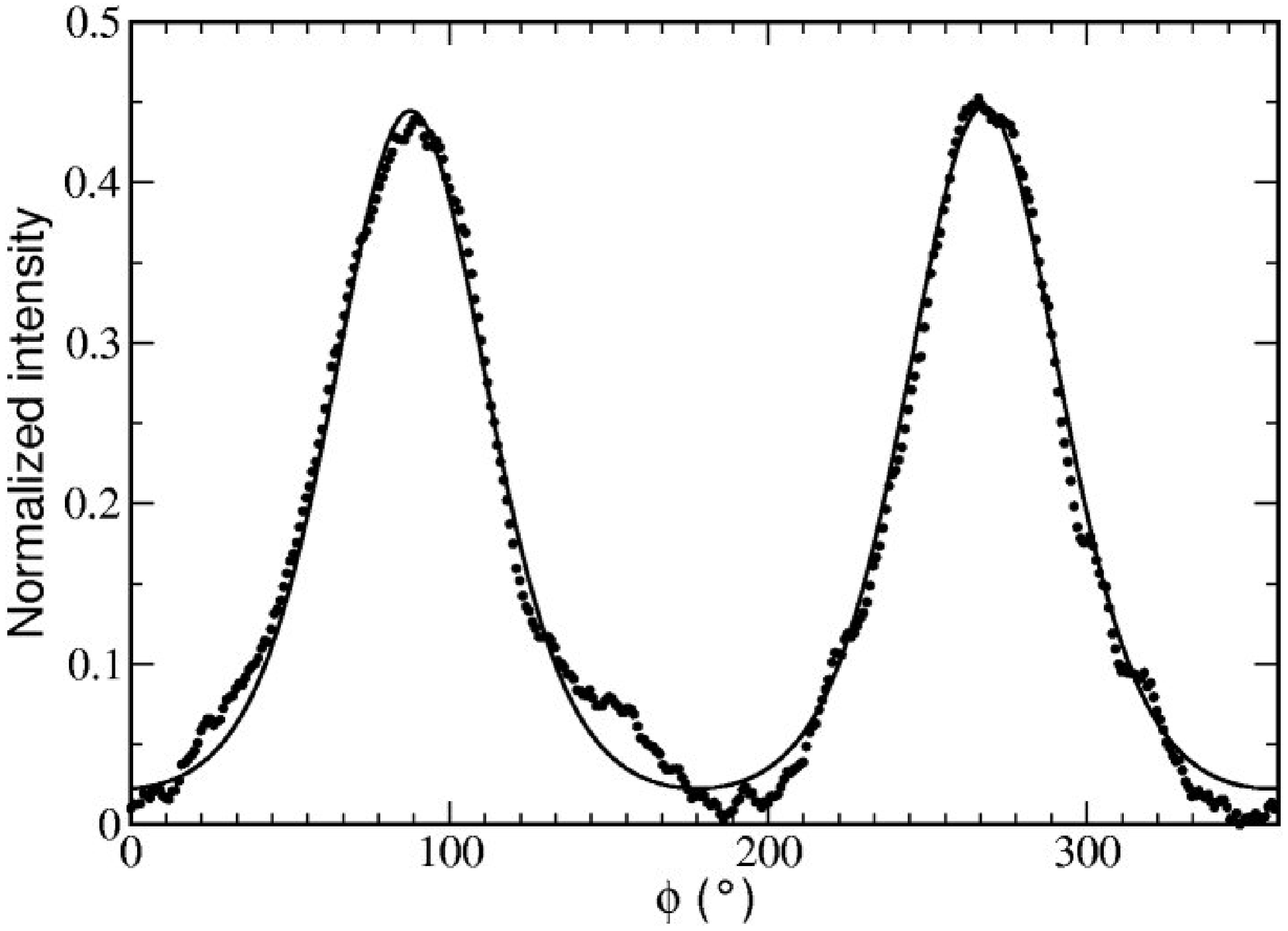}
\end{figure}



\end{document}